\documentclass[acus]{Jac2000}

\usepackage{graphicx}

\begin{document}
\title{MEASURING AND CONTROLLING ENERGY SPREAD IN CEBAF\thanks{
   Work supported by the United States Department of Energy under Contract
No.~DE-AC05-84ER40150.}}

\author{G.~A.~Krafft, J.-C.~Denard, R.~W.~Dickson, R.~Kazimi,\\ V.~A.~Lebedev,
and M.~G.~Tiefenback\\ TJNAF, Newport News, VA23606, USA}

\maketitle

\begin{abstract} 
     As compared to electron storage rings, one advantage of recirculating
linacs is that the beam properties at target are no longer dominated by the
equilibrium between quantum radiative diffusion and radiation damping because
new beam is continually injected into the accelerator. This allows the energy
spread from a CEBAF-type machine to be relatively small; the measured energy
spread from CEBAF at 4 GeV is less than 100 parts per million accumulated over
times of order several days. In this paper, the various subsystems contributing
to the energy spread of a CEBAF-type accelerator are reviewed, as well as the
machine diagnostics and controls that are used in CEBAF to ensure that a small
energy spread is provided during routine running. Examples of relevant
developments are (1) stable short bunches emerging from the injector, (2)
precision timing and phasing of the linacs with respect to the centroid of the
beam bunches on all passes, (3) implementing 2 kHz sampling rate feedback
systems for final energy stabilization, and (4) continuous beam energy spread
monitoring with optical transition radiation devices. We present measurement
results showing that small energy spreads are achieved over extended periods.

\end{abstract}

\begin{figure}[htb]
\centering
\includegraphics*[width=92mm]{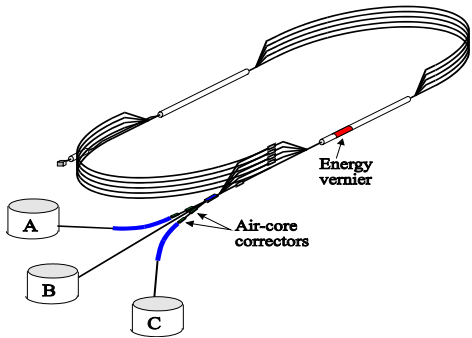}
\caption{Schematic of the CEBAF Accelerator}
\label{cebafacc}
\end{figure}

\section{INTRODUCTION}

In this paper we summarize the present status on energy spread
measurement and control in the Jefferson Lab nuclear physics accelerator
called CEBAF. A schematic diagram of CEBAF appears in Fig.~\ref{cebafacc}
and a summary of relevant beam parameters is given in Table~\ref{cebafpar},
where all sizes are {\it rms} quantities.
CW beam, originating in the injector, is recirculated up to five times through
each linac. The beam may be directed into up to three experimental halls
simultaneously, the beam current in the halls being at the third subharmonic of
the
accelerator operating frequency of 1497 MHz. Because of the low
charge-per-bunch at even the highest operating current,
collective effects are not an important source of energy spread in CEBAF.
Fig.~\ref{cebafacc} locates some of the feedback system hardware, discussed
in Section
6 below.

\begin{table}[htb]
\begin{center}
\caption{CEBAF Accelerator Parameters}
\begin{tabular}{|l|c|c|}
\hline
\textbf{Item} & \textbf{Value} & \textbf{Unit} \\ \hline
Beam Energy& 0.8-6 & GeV \\
Beam Current& 180 & $\mu$A/Hall \\
Normalized {\it rms} Emittance& 1 & mm mrad \\
Repetition Rate & 499 & MHz/Hall \\
Charge per Bunch&$<$0.4&pC\\
Extracted {\it rms} Energy Spread&$<10^{-4}$&\\
Transverse {\it rms} Beam Size&$<$100&$\mu$m\\
Longitudinal {\it rms} Beam Size&60(200)&$\mu$m(fsec)\\
Beam {\it rms} Angular Spread&$<0.1/\gamma$&\\ \hline
\end{tabular}
\label{cebafpar}
\end{center}
\end{table}

\section{SOURCES AND TYPES OF ENERGY SPREADS IN RECIRCULATING LINACS}

Because the electron beam remains in CEBAF for
times that are short compared to the usual radiation damping times for the
recirculation rings, the energy spread of the recirculating beam is not
determined by the equilibrium defined by the quantum
character of the emission of synchrotron radiation. What
effects do determine the energy spread? The sources of energy
spread will be grouped into two broad categories: the single bunch energy
spread
which is the
same for all bunches in the train, and fluctuation energy spread which is
derived from fluctuations in the beam centroid energy.

Sources of single bunch energy spread are: (1) the injected
single bunch
energy spread, (2) energy spread generated by the finite phase extent of
the bunches interacting with the time-dependent accelerating field, (3)
synchrotron emission in the arcs, (4) average phase errors in the synchronization of
the cavity RF to the beam, (5) summed phase errors from whole linac sections
that are not properly balanced, and (6) interactions of the beam energy spread
with non-zero $M_{56}$ in the arcs, which might cause the injected bunch length
to grow.

Sources of energy fluctuations are: (1) RF phase
 fluctuations in individual RF cavities, (2) RF amplitude errors in the
individual cavities, (3) master oscillator noise, and (4) magnetic field
fluctuations in dipole magnets that are used for energy measurements
by the feedback system. We
will address each of these potential sources of energy spread. The general
philosophy used at CEBAF is to use measurements to ensure the machine setup
minimizes
the single bunch energy spread, and to use feedback systems to correct
energy fluctuations. Our point-of-delivery diagnostics
allow us to ensure that the energy spread is under control throughout the
duration of physics running.

\section{LONGITUDINAL MANIPULATIONS IN INJECTOR}

In general terms, the function of the injector is to accelerate the
electron beam to an energy high enough that the phase slip
caused by different
passes being at different energies is small, and to manipulate the
longitudinal phase space of the beam
in a way that minimizes the overall extracted
energy spread. To solve the first problem, 45 MeV injection energy is
sufficient for 4 GeV total acceleration. A simple calculation gives guidance
 on injection conditions that produce the optimal energy spread. Assume for
 the moment that one
could phase each linac cavity for exactly maximum energy gain. Then the
energy of a bunch electron after leaving the accelerator is
$E=E_{inj}+E_{gain}\cos(\Phi)$
where $E_{inj}$ and $E$ are the initial and final energy, respectively,
and $\Phi$ is the phase of the electron with respect to the bunch centroid
(assumed on crest at $\Phi=0$). Utilizing the single particle distribution
function for the electrons at injection to perform the proper statistical
averages one obtains
$$\sigma_E/E=\sqrt{\sigma_{E,inj}^2/E^2+\sigma_\Phi^4/2},$$
where $\sigma_E$ and $\sigma_{E,inj}$ are the {\it rms} energy spreads after
acceleration and at injection, respectively, and $\sigma_\Phi$ is the
{\it rms} phase spread at injection. The first term damps as energy is
increased because the initial spread becomes a smaller part of the total,
whereas the final term does not depend on the energy because both energy
and energy spread accumulate at
the same rate due to a non-zero bunch length.
Given a certain longitudinal emittance from the source $\epsilon_l$ and the
final energy, there is an optimum energy spread of $\sigma_E/E=
\sqrt{3/2}(\epsilon_l/E)^{2/3}$ at the optimal injected bunch length of
$\sigma_{\Phi,opt}=(\epsilon_l/E)^{1/3}$.
A typical measured value for $\epsilon_l$ is 6.7 keV $^\circ$, yielding an
optimal energy spread of $1.16\times 10^{-5}$ at 4 GeV, with a bunch length
of $\sigma_\Phi=0.18^\circ= 320$ fsec.
A primary function of the injector is to
provide a longitudinal phase space ``matched" to this bunch length.

A way of providing this match has been developed and documented in various
conference proceedings and workshops \cite{kaz2,kr7}.
Here
we concentrate on the measurements done
 routinely
to ensure that proper bunching has been achieved. A main diagnostic used
at CEBAF
is to perform phase transfer function measurements \cite{yao1,kr5}.
The basic idea is to phase modulate the beam
at the beam chopper, with the rest of the RF phases in the accelerator
held constant. By analyzing the longitudinal transfer function for
its linear and
non-linear behavior, one has a way to ensure that the beam longitudinal
phase space is bunched in a way that minimizes distortion in the bunching
process, 
including the
non-linearities due to RF curvature and higher order terms \cite{kr7}.
Such measurements
are routinely used to restore the proper operation of the injector after
machine downs, or when certain types of
operational problems arise.

\begin{figure}[htb]
\centering
\includegraphics*[width=85mm]{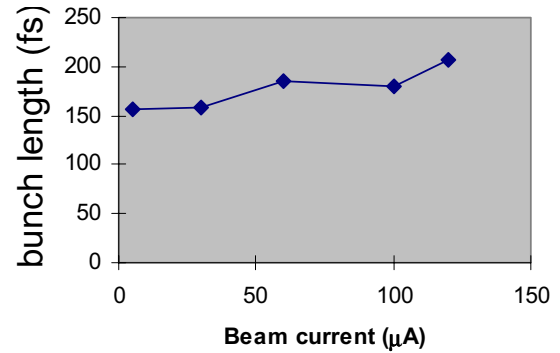}
\caption{Bunch Length vs.~Beam Current out of Injector}
\label{kazimi}
\end{figure}

Next we present a summary viewgraph from another talk at
this conference, which shows that the bunch length is properly adjusted
\cite{kaz1}.
In
Fig.~\ref{kazimi}, we present the bunch length as measured by the
zero-phasing method \cite{wang}, as a function of current over
the full operating range of the CEBAF accelerator. One observes a roughly
constant bunch length between 150 and 200 fsec (45 and 60 $\mu$m). This value
is matched well enough that the extracted single bunch energy spread
is less than $1.5\times 10^{-5}$.

What are the effects associated with
breaking the assumption of ideal phasing? In an analysis that was used
to set tolerances for the RF controls \cite{kr3,k4,sim1}, it was demonstrated
that
as long as: (1) the uncorrelated amplitude errors in the cavities were
under $2\times10^{-4}$, (2) the uncorrelated phase fluctuations in the
cavities were under several tenths of a degree, (3) the phasor sum
of the gradients obtained from each cavity is purely real, and
(4) the thermal drifts along the linac were stabilized to an error less
than 2.6$^\circ$, then the resulting energy fluctuations in the beam would
be less than $2.5\times10^{-5}$ for an assumed bunch length of 0.3$^\circ$.
Another way of stating condition
(3) is that for each pass through the accelerator, one would like to
arrive at the time that provides the crest energy for the whole linac.
Next, we discuss how this condition is achieved in practice.

\section{PATH LENGTH AND $M_{56}$}

Suppose for the moment that the phase of one pass through one linac was
off crest by $\Phi_e$ radians. Then the relative energy spread
generated
by this error is $\sigma_e/E=\sigma_\Phi\Phi_e/10$, the factor of ten
appearing because we have assumed one linac pass is not phased properly out
of ten linac
passes total. To have the resulting energy spread at 10 ppm, one
needs the phase error to be less than $35$ mrad = 2$^\circ$ for a bunch length
of 300 fsec.

Likewise, suppose that we require less than 10\% growth in the bunch length
going through each arc of CEBAF. By a statistical argument, there will
be less than
30\% bunch length growth after going through the nine arcs of the CEBAF
accelerator. Given a beam energy spread less than 10$^{-4}$, the $M_{56}$ of the
arcs should not exceed 10 cm, a fairly weak limitation.

Presently, the apparatus in routine use to perform this measurement is based
on measuring the time-of-arrival of each separate beam pass with a longitudinal
pickup cavity tuned to the beam fundamental, whose output is mixed with
the master
oscillator in a phase detector arrangement \cite{c1}.
The development of this device
from first experiments to final instrument is documented in several Particle
Accelerator Conference contributions \cite{c1,k1,h2}. Because only relative
times-of-arrival
are required, the precision of the method is very high. With 4.2 $\mu$sec
4 $\mu$A beam pulses, a precision of 0.1$^\circ$ = 185 fsec is routinely
achieved. Such precision is clearly sufficient for setting the path length,
and allows $M_{56}$ of the arcs to be determined to under 3 cm by an energy
modulation experiment where the energy is changed by $2\times10^{-3}$.

\section{MASTER OSCILLATOR MODULATION SYSTEM}

There is a significant limitation in the present system used to set the
path length. Path length checks must be done
invasively to normal beam delivery, by going into a pulsed beam mode.
It would be far better to have a method to monitor
the linac phases, including higher passes, continuously and accurately.
During the last few years a system has been developed that will allow
continuous monitoring and cresting of the linacs on all passes \cite{tief1}.
This
system had its origin in an automatic beam-based linac cresting routine
\cite{tief2},
and it
is already used routinely to set the first pass through
each linac close to crest.

The system takes advantage of the CW electron beam delivered by CEBAF
and standard lockin techniques. It is based on phase modulating
the master reference going to each of the linacs, at 383 Hz for measurements
of the
first, so-called north linac, and at 397 Hz for the south linac.
Simultaneously
and coherently with the modulations, one observes the position motion on
a beam position monitor (BPM) downstream of both linacs at a point of non-zero
dispersion. Linac cresting corresponds to zero output from the BPM at
the modulation frequency. Long integration times permit cresting to
be performed with high precision.
The required
phase modulation is small enough that the energy spread generated by the
dither remains small.

Table~\ref{momod} summarizes the system parameters and performance of
the Master
Oscillator Modulation system. Its performance, especially in the next step
in setting the
higher pass beams close to crest, should allow us to reduce the energy
spread of the extracted beam by roughly a factor of two.

\begin{table}[htb]
\begin{center}
\caption{Master Oscillator Modulation System}
\begin{tabular}{|l|c|c|}
\hline
\textbf{Item} & \textbf{Value} & \textbf{Unit} \\ \hline
Modulation Amplitude & 0.05 & 1497 MHz $^\circ$ \\
Modulation Frequencies & 383, 397 & Hz \\
Sensitivity & $>$6000 & $\mu V/^\circ$ \\
Operating Current & $>$2 & $\mu$A \\
Dispersion at BPM & 1.4 & m \\
Measurement Precision& $<$0.1 & 1497 MHz $^\circ$ \\ \hline
\end{tabular}
\label{momod}
\end{center}
\end{table}
\section{FAST FEEDBACK SYSTEM}

As mentioned in the introduction, sources of bunch energy centroid
fluctuations are corrected by a fast digital feedback system
\cite{dickson}.
The system corrects beam position and energy near the targets of the
nuclear physics experiments utilizing energy measurements obtained from
the bend magnets which deliver the beam to the various halls, see Fig.~\ref
{cebafacc}. The system is capable of suppressing beam motion in the
frequency band from 0 to 80 Hz and also performs narrow band suppression
at the first twelve power line harmonics. The system operates with a 2.1 kHz
sampling rate and utilizes two VME board computers to compute the corrections.
Energy
corrections are fed back as analogue signals
to the gradient set points
in the RF controls of a few cavities in the linac called vernier cavities.

For the standard optics in CEBAF, the horizontal dispersion is maximum in
the middle of the bend magnets delivering the beam to the halls. Its value
is approximately 4 m, meaning position fluctuations at 10 ppm correspond to
40 $\mu$m of beam motion. The feedback system suppresses the fluctuations
to around 20 $\mu$m, limited by BPM noise \cite{dickson}. The beam noise
to be corrected is primarily at frequencies of 60 Hz and its first few
harmonics.

Because the energy information is so closely tied to the magnetic
fields in the beam delivery lines to the halls, a question arises about
the stability of the magnetic fields themselves at
the 10$^{-5}$ level. The total
magnetic
field at several points within the magnets have been verified to be stable to
10 ppm, and the power supplies deliver current having
similarly small fluctuations.
Recently, we
have installed a
magnetic flux loop monitor through the dipole strings to
the halls. This monitor will provide better quantitative information than
we currently possess on residual fluctuations in the magnets,
and will be able to address the issue of magnet stability directly.

\section{ENERGY SPREAD DIAGNOSTICS}

The accelerator is equipped with slow wire scanners using 22 $\mu$m diameter
tungsten
wires. Beam profiles for currents in the 2 to 5 $\mu$A range can be accurately
measured once a minute with such scanners. More recently, we have
developed
a profile monitor that can measure even
the most intense beams using forward optical
transition radiation (OTR) \cite{p1, d1}. A very thin (1/4 $\mu$m) carbon foil
inserted
into the beam path is not invasive to physics experiments for most
CEBAF energies and currents. Presently, OTR monitors are installed in
each of the Experimental Hall A
and C beam transport lines at the high dispersion points
of the beam optics.
These monitors
provide
the experiments and the accelerator with 5 Hz measurement rates for each
instrument by using a common image processing hardware. A dedicated software,
developed under the EPICS \cite{delas}
control system, multiplexes up to four video input
channels connected to a single MaxVideo image processing board \cite{h1}. The
global
processing speed is 10 Hz, 5 Hz for each of the two OTR
monitors. The Hall A OTR measures beams in the 1 to 180 $\mu$A operational
range;
in Hall C, the dynamic range extends down to 0.1 $\mu$A. The resolution of these
monitors is limited by the CCD camera to about 2 pixels. This amounts to
approximately 70 $\mu$m of {\it rms} beam size.

\begin{figure}[htb]
\centering
\includegraphics*[width=92mm]{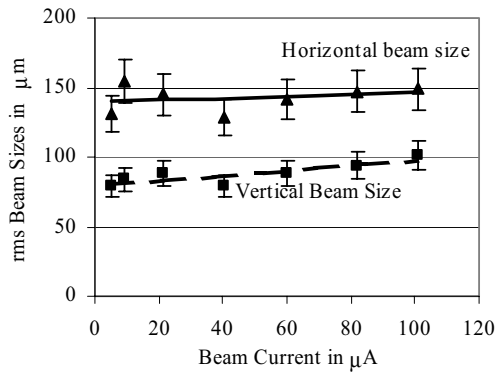}
\caption{
Beam Size vs.~Beam Current at high dispersion point in beam delivery line.}
\label{li2000f3}
\end{figure}

Fig.~\ref{li2000f3}
shows that the energy spread is relatively stable and below
$ 4\times 10^{-5}$
for a wide range of beam currents.
The horizontal size, measured at the 4 m dispersion point, is mostly due to
the energy spread. Neglecting the betatron beam size, 40 $\mu$m, and the camera
resolution, 70 $\mu$m, overestimates the actual energy spread by less than
25\%.

Continuous small energy spread became an operational requirement
at CEBAF in
Dec.~1999, for a hypernuclear experiment housed in Hall A,
and continued
until May 2000 with a similar experiment in Hall C. Both
experiments
ran simultaneously during one month last March,
with 2-pass
beam for Hall A and a 4-pass beam for Hall C.
Delivering two beams with
tight energy spread and energy stability requirements instead of one proved
demanding.
The energy requirements for each experiment were similar: $dp/p \le 5\times
10^{-5}$,
with energy stability better than $1\times 10^{-4}$. In addition,
Hall A needed the
transverse beam sizes at the target to be less than 200 $\mu$m but
greater than 100 $\mu$m and a beam position stable within 250 $\mu$m.

\begin{figure}[htb]
\centering
\includegraphics*[width=92mm]{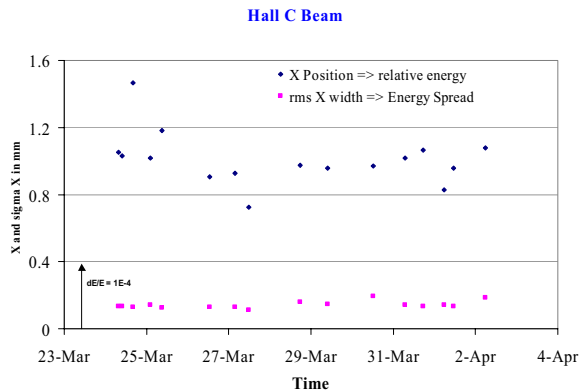}
\caption{Horizontal position and size
of Hall C beam during delivery period. Note that
the Hall C beam line provided the energy locking data.}
\label{li2000f4}
\end{figure}

\begin{figure}[htb]
\centering
\includegraphics*[width=92mm]{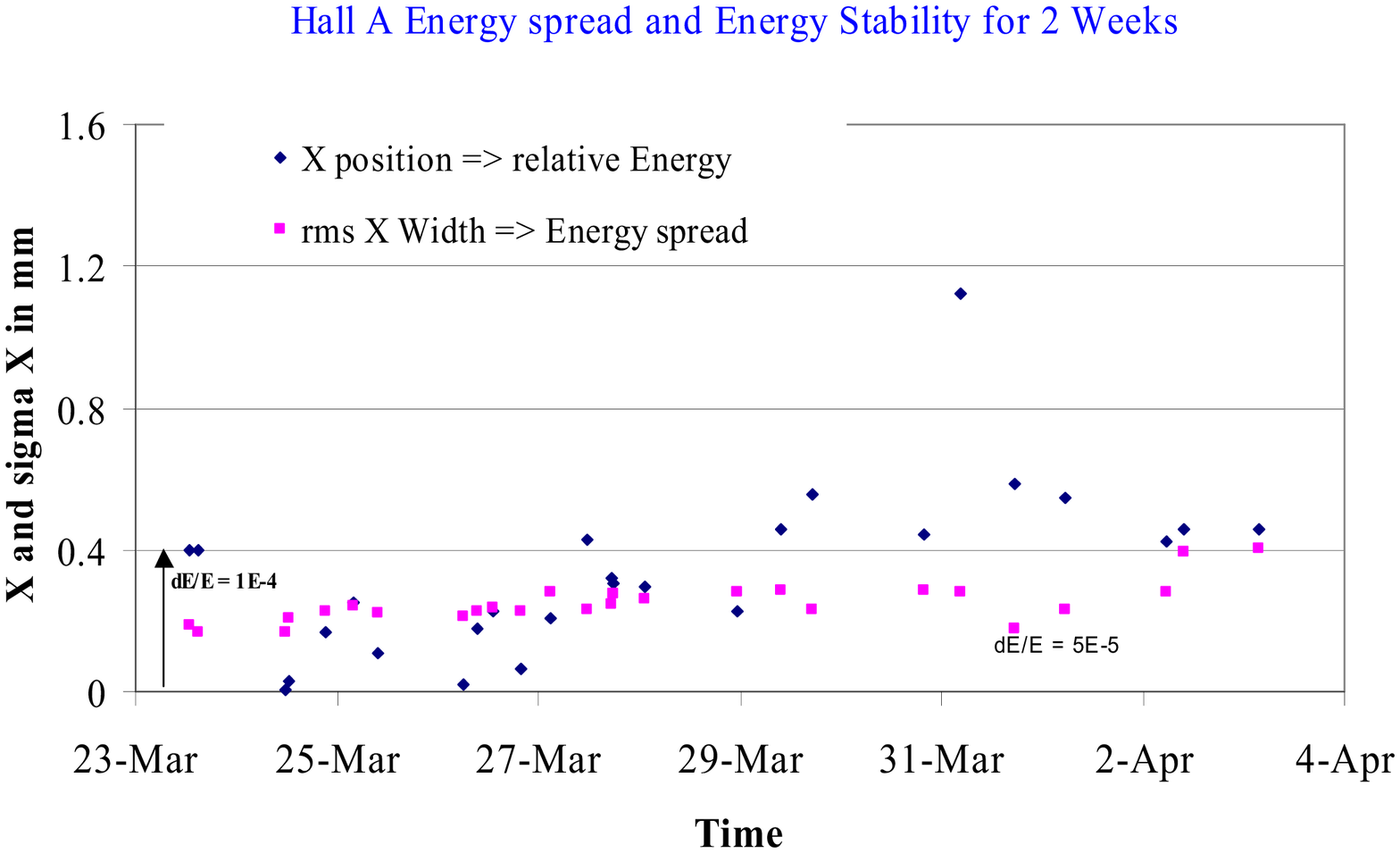}
\caption{Horizontal position and size
of Hall A beam during delivery period. Note the
degradation of the spread with time due to uncorrected drifts.
Even with drifts, the spread is remarkably small.}
\label{li2000f4i}
\end{figure}

The energy spread requirements have been routinely achieved
for the hall under feedback control. Because the feedback system
can correct the energy fluctuations only in a single hall as presently
configured, there were uncertainties that the spread in the other
hall would remain small.
Fig.~\ref{li2000f4}
shows energy spreads and
relative energies in the Hall C beam
recorded over a 2-week period, with Hall C transport line
providing the energy corrections to the energy feedback system.
Small energy spreads were delivered throughout the period to
Hall C, however drifts led to energy spread increases in the Hall A beam,
as seen in Fig.~\ref{li2000f4i}.

Throughout the experiments in either hall, the energy spread and stability
of both beams were continuously
recorded. The OTR monitors have been critical in this task. They were initially
too cumbersome to be easily used by all operator crews. The implementation of
scripts that periodically check and adjust the camera illumination, that
initialize the image processing board according to the beam, and that set a
``data valid" flag quickly improved the instrument availability to 95\%
\cite{h1}. After these improvements,
the machine crews were able to correct quickly unacceptable energy spreads,
usually without interrupting beam delivery.

 We are planning to improve the
energy spread monitoring for two reasons: 1) At lower energies ($<$1.2
GeV), the beam current had to be lowered to under 50 $\mu$A to have acceptable
radiation levels on sensitive beam-line equipment. 2) Experiments scheduled
in 2002 require monitoring an energy spread of $2\times10^{-5}$.
As an alternate to OTR monitoring,
we are planning to use synchrotron light beam monitoring,
which is
less invasive to the experimenters. However, the resolution of such a device is
limited to about 100 $\mu$m in the visible using the bending magnets of the
hall transport lines. We are starting a development effort in order to reach
about 30 $\mu$m resolution utilizing the UV synchrotron emission.

\section{CONCLUSIONS}
We have demonstrated the ability of a CEBAF-type accelerator to produce beams
with small energy spreads over long periods of time. We ensure that the energy
spread remains small by: (1) ensuring the bunch length out of the injector is
small,
(2) ensuring that the beam remains close to the crest phase on each separate
pass
(soon continuously and automatically!), and (3) providing
continuous fast correction of
60 cycle harmonic noise on the beam. We have developed beam
diagnostic devices to continuously monitor and record beam conditions with
5 Hz update rates using digitization of multiple video monitors.

\end{document}